\documentclass{emulateapj}
\usepackage{enumitem}

\newcommand{\kms}{\ifmmode {\rm km~s}^{-1} \else km~s$^{-1}$\fi}
\newcommand{\Msun}{\ifmmode {\rm M}_{\odot} \else M$_{\odot}$\fi}
\newcommand{\Lsun}{\ifmmode {\rm L}_{\odot} \else L$_{\odot}$\fi}
\newcommand{\qo}{\ifmmode q_{\rm o} \else $q_{\rm o}$\fi}
\newcommand{\Ho}{\ifmmode H_{\rm o} \else $H_{\rm o}$\fi}
\newcommand{\ho}{\ifmmode h_{\rm o} \else $h_{\rm o}$\fi}

\newcommand{\vFWHM}{\ifmmode v_{\mbox{\tiny FWHM}} \else
                    $v_{\mbox{\tiny FWHM}}$\fi}
\newcommand{\CCF}{\ifmmode F_{\it CCF} \else $F_{\it CCF}$\fi}
\newcommand{\ACF}{\ifmmode F_{\it ACF} \else $F_{\it ACF}$\fi}
\newcommand{\Halpha}{\ifmmode {\rm H}\alpha \else H$\alpha$\fi}
\newcommand{\Hbeta}{\ifmmode {\rm H}\beta \else H$\beta$\fi}
\newcommand{\Hgamma}{\ifmmode {\rm H}\gamma \else H$\gamma$\fi}
\newcommand{\Hdelta}{\ifmmode {\rm H}\delta \else H$\delta$\fi}
\newcommand{\Lya}{\ifmmode {\rm Ly}\alpha \else Ly$\alpha$\fi}
\newcommand{\Lyb}{\ifmmode {\rm Ly}\beta \else Ly$\beta$\fi}
\newcommand{\HeI}{\ifmmode {\rm He}\,{\sc i}\,\lambda5876 \else 
	          He\,{\sc i}\,$\lambda5876$\fi}
\newcommand{\HeII}{\ifmmode {\rm He}\,{\sc ii}\,\lambda4686 \else 
	           He\,{\sc ii}\,$\lambda4686$\fi}

\newcommand{\heii}{He\,{\sc ii}}

\newcommand{\ciii}{\ifmmode {\rm C}\,{\sc iii} \else C\,{\sc iii}\fi}
\newcommand{\civ}{\ifmmode {\rm C}\,{\sc iv} \else C\,{\sc iv}\fi}
\newcommand{\CIV}{\ifmmode {\rm C}\,{\sc iv}\,\lambda1549 \else 
	           C\,{\sc iv}\,$\lambda1549$\fi}

\newcommand{\niv}{N\,{\sc iv}]}

\newcommand{\mgii}{Mg\,{\sc ii}}

\shorttitle{Biases Discovered due to Low S/N Spectra}
\shortauthors{}

\received{}

\begin{document}

\title{The Sloan Digital Sky Survey Reverberation Mapping Project: An Investigation of Biases in \civ\ Emission-Line Properties}

\author{ K.~D.~Denney\altaffilmark{1,2,3}, Keith~Horne\altaffilmark{4}, W.~N.~Brandt\altaffilmark{5,6,7}, Luis~C.~Ho\altaffilmark{8,9}, B.~M.~Peterson\altaffilmark{1,2}, Gordon~T.~Richards\altaffilmark{10}, Yue Shen\altaffilmark{11,12}, J.~R.~Trump\altaffilmark{5,13}, J.~Ge\altaffilmark{14}}

\altaffiltext{1}{Department of Astronomy, 
		The Ohio State University, 
		140 West 18th Avenue, 
		Columbus, OH 43210, USA;
		denney@astronomy.ohio-state.edu}
		
\altaffiltext{2}{Center for Cosmology and AstroParticle Physics, 
                 The Ohio State University,
		 191 West Woodruff Avenue, 
		 Columbus, OH 43210, USA}

\altaffiltext{3}{NSF Astronomy \& Astrophysics Postdoctoral Fellow}

\altaffiltext{4}{SUPA Physics/Astronomy, 
		    Univ. of St. Andrews, 
		    St. Andrews KY16 9SS, Scotland, UK}

\altaffiltext{5}{Department of Astronomy \& Astrophysics, 
                      525 Davey Lab, The Pennsylvania State University, 
                      University Park, PA 16802, USA}
			
\altaffiltext{6}{Institute for Gravitation and the Cosmos, 
                      The Pennsylvania State University, 
                      University Park, PA 16802, USA}
			
\altaffiltext{7}{Department of Physics, 
                      104 Davey Lab, The Pennsylvania State University, 
                      University Park, PA 16802, USA}

\altaffiltext{8}{Kavli Institute for Astronomy and Astrophysics, 
                     Peking University, 
                     Beijing 100871, China}

\altaffiltext{9}{Department of Astronomy, 
                      School of Physics, 
                      Peking University, 
                      Beijing 100871, China}

\altaffiltext{10}{Department of Physics, 
		     Drexel University, 3141 Chestnut St., 
		     Philadelphia, PA 19104, USA}

\altaffiltext{11}{Department of Astronomy, University of Illinois at Urbana-Champaign, Urbana, IL 61801, USA}

\altaffiltext{12}{National Center for Supercomputing Applications, University of Illinois at Urbana-Champaign, Urbana, IL 61801, USA}

\altaffiltext{13}{Hubble Fellow}
		    
\altaffiltext{14}{Astronomy Department
			University of Florida
			211 Bryant Space Science Center
			P.O. Box 112055
			Gainesville, FL 32611-2055, USA}


\begin{abstract}
  We investigate the dependence on data quality of quasar properties measured from the \civ\ emission line region at high redshifts.  Our measurements come from 32 epochs of Sloan Digital Sky Survey (SDSS) Reverberation Mapping Project spectroscopic observations of 482 $z>1.46$ quasars.  We compare the differences between measurements made from the single-epoch and coadded spectra, focusing on the \CIV\ emission line because of its importance for studies of high-redshift quasar demographics and physical properties, including black hole masses.  In addition to increasing statistical errors (by factors of $\sim$2$-$4), we find increasing systematic offsets with decreasing S/N.  The systematic difference (measurement uncertainty) in our lowest S/N ($<$5) subsample between the single-epoch and coadded spectrum (i) \civ\ equivalent width is 17\AA\ (31\AA), (ii) centroid wavelength is $<$1\AA\ (2\AA), and fractional velocity widths, $\Delta V/V$, characterized by (iii) the line dispersion, $\sigma_l$, is 0.104 (0.12), and (iv) the mean absolute deviation (MAD) is 0.072 (0.11). These remain smaller than the 1$\sigma$ measurement uncertainties for all subsamples considered. The MAD is found to be the most robust line-width characterization. Offsets in the \civ\ full-width at half maximum (FWHM) velocity width and the \civ\ profile characterized by FWHM/$\sigma_l$ are only smaller than the statistical uncertainties when S/N$>$10, although offsets in lower S/N spectra exceed the statistical uncertainties by only a factor of $\sim$1.5.  Characterizing the \civ\ line profile by the kurtosis is the least robust property investigated, as the median systematic coadded--single-epoch measurement differences are larger than the statistical uncertainties for all S/N subsamples.

\end{abstract}




\section{INTRODUCTION}

Since their discovery \citep{Schmidt63}, quasars (synonymously referred to in this work as active galactic nuclei or AGN) have evolved from a curiosity to a powerful cosmological probe.  They are now understood to be the visible growth phase of the supermassive black holes (BHs) that reside at the center of massive galaxies.  The strong correlations between the properties of these BHs and their host galaxies \citep[][see also the review by \citealt{Kormendy&Ho13} and references therein]{Kormendy&Richstone95, Magorrian98, Ferrarese00, Gebhardt00b, Gebhardt00a, Gebhardt00b, Ferrarese01,Tremaine02, Wandel02, Onken04, Nelson04, Graham07, Bentz09ml, McConnell13} strongly suggest that BHs may play an active role in the evolution of galaxies and their surroundings, presumably due to feedback produced from the energy released by the gravitational accretion of material onto the BH \citep[e.g.,][]{Hopkins&Elvis10, Debuhr11, Fabian12}. Whether the growth and co-evolution of BHs and galaxies is mutually regulating and causal or not \citep{Jahnke&Maccio10, Sun15a}, these systems remain important probes of the distant and nearby universe.

Many physical properties of the accreting BHs, such as the mass and accretion rates, can be estimated from a single quasar spectrum \citep[e.g.,][]{Mclure02, Vestergaard02, Vestergaard06, Wang09, Rafiee11, Park13}.  This is possible through largely empirical scaling relationships calibrated using small samples of low-redshift AGN.  In particular, reverberation mapping \citep{Blandford82, Peterson93, Peterson14} provides the calibration for estimating ``single-epoch" (SE) quasar masses, using only two observables measurable from a single quasar spectrum:  the broad line region (BLR) velocity --- inferred from the velocity-width of a broad emission line --- and the radius of the variable BLR gas --- inferred from the quasar luminosity \citep{Laor98, Wandel99, Kaspi00, Bentz09rl, Bentz13}.  

Using quasars for studying cosmology, galaxy evolution, and BH accretion physics requires an understanding of their evolution in number and mass over cosmic time.  This requires measuring these properties for large samples across the universe.  In the last several decades, photometric and spectroscopic quasar surveys operating over a wide range of wavelengths and energies have vastly increased the number of known quasars and their redshifts.  In particular, the Sloan Digital Sky Survey \citep[SDSS;][]{York00} has cataloged $\sim$300,000 spectroscopically confirmed quasars when combining the fifth edition SDSS Quasar Catalog \citep{Schneider10} with the Data Release 10 Quasar catalog \citep[DR10Q;][]{Paris14} from the Baryon Oscillation Spectroscopic Survey \citep[BOSS;][]{Dawson13}.  

Large statistical samples can enable detailed studies of quasar demographics only if the systematic biases associated with survey limits and data quality are well understood. In this study we investigate possible biases in several spectroscopic properties derived for high-redshift quasars due to the low signal-to-noise ratio (S/N) of typical survey spectra.  Since the goal of most surveys is to measure redshifts with which to map the universe, this only requires a high enough S/N to reliably detect a high-equivalent-width emission line.  Despite the intrinsic high luminosity of quasars, the vast majority will have observed fluxes close to the flux limit of the survey \citep[e.g.,][]{Paris14}.  As a result, the quasar continuum and lower equivalent width features (such as those used for BH masses and reliable redshifts) will be at significantly lower S/N.  Consequently, it is important to understand the data-quality limit at which the reliability (defined by both precision and accuracy) of an investigation becomes compromised. 

The data used for direct, reverberation-mapping based BH measurements are generally of very high quality.  This is also true of the data used to calibrate SE virial BH mass scaling relations \citep[e.g.,][]{Park13}.  As such, our understanding of the general uncertainty in BH masses derived either directly through reverberation mapping or indirectly through empirical scaling relationships is based on high-quality data. Yet, most of the higher-redshift quasar spectra to which these scaling relations are applied are typically low-S/N, ``survey-quality"\footnote{Here, we use ``survey-quality'' to refer to the typically lower spectral S/N ($\lesssim$3$-$5 per pixel at the flux limit) data that is often the product of large, moderate-resolution, flux-limited, redshift surveys, such as the SDSS.} data.  There can be systematic differences between spectroscopic properties measured from the high-quality calibration data and the survey-quality data, leading to systematic errors in BH mass estimates.  \citet{Denney09a} examined this for \Hbeta\ line width measurements for two low-redshift, relatively low-luminosity AGN, and found statistical and systematic effects that became a significant source of bias for data with spectral S/N$<$10 (per pixel) in the continuum.  

Here, we investigate the effects of low S/N on \civ\ emission-line properties.  We first evaluate statistical and systematic errors in the estimates of the velocity width of the broad \CIV\ emission line. The broad emission line velocity width is one of the two observables needed to estimate a virial BH mass, which allows subsequent studies of cosmic structure growth, seed black holes, and the co-evolution of galaxies and BHs.  We also look at other commonly-measured \civ\ emission-line properties, including the \civ\ equivalent width, centroid, and line shape, that may be useful in determining further details of the accretion and feedback processes, structure, and kinematics of the central engine and/or for evaluating other biases in \civ-based black hole masses \citep[e.g.,][]{Denney12}.

These measurable spectroscopic properties are important for direct studies of quasar physics and demographics, as well as the broader evolutionary studies based thereon, as the \civ\ emission line is present in the optical wavelength regime in the range $1.4\lesssim\, z\lesssim\,4.8$, which covers the quasar epoch and extends to a time when the universe was less than a tenth of its current age. In Section \ref{S_Data} we present the data we use for this investigation, while Section \ref{S_SpecAnal} describes the spectroscopic analysis used to fit and measure the properties of interest from our quasar sample.  We discuss trends and other results as a function of the data quality and other quasar properties in Section \ref{S_Discussion} and make concluding remarks about our results in Section \ref{S_Conclusions}. 

\section{\civ\ Spectral Data}
\label{S_Data}

In 2014, the SDSS Reverberation Mapping Project (SDSS-RM) spectroscopically monitored 849 broad-line quasars in the CFHT-LS W3 field (which is also the AEGIS field and a PanSTARRS medium-deep field) with the BOSS spectrograph \citep{Smee13} that is mounted on the SDSS telescope \citep{Gunn06} as part of an ancillary program of the SDSS-III surveys \citep{Eisenstein11}.  The SDSS-RM sample covers redshifts over the range $0.1 < z < 4.5$ down to a flux limit of $i_{\rm psf} = 21.7$ mag in a single 7 deg$^2$ field.  Each of the 32 epochs of observations was a $\sim$2 hr exposure taken during dark/grey time with an average cadence of $\sim$4 days over a period of 6 months.  Further details of the SDSS-RM project, the technical overview, and program goals are provided by \citet{Shen15b}.  

\begin{figure}
\epsscale{1.2}
\plotone{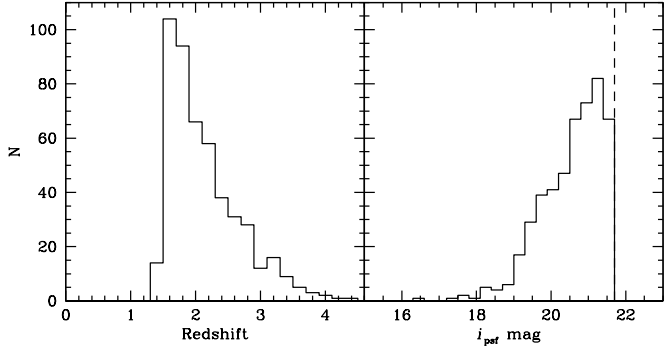}

\caption{Distribution of redshifts and $i_{\rm psf}$ magnitudes for the sample of 482 SDSS-RM quasars.  The vertical dashed line in the right panel shows the magnitude limit for the SDSS-RM sample.} 

\label{F_physprop}
\end{figure}

Here, we only consider objects from the SDSS-RM sample with $z>1.46$ so that the \civ\ emission line and a short ward continuum region are accessible.  We further removed three objects for which low-EW emission lines, although clearly seen in the coadded spectrum, were not easily discernible in all or most of the single-epoch spectra, leading the fits (described below) to fail or provide bogus results for most SE spectra.  We also removed 12 broad absorption line (BAL) quasars in which the BAL absorbed too much of the \civ\ emission line for the width to be reliably characterized.  On occasion, objects were discovered to have ``dropped spectra", where the fiber was not properly plugged into the plate. We visually inspected all epochs for all 482 quasars, removing any epoch in which the spectrum was completely absent due to this or other problems.  We did not initially drop any spectra based on their S/N.\footnote{Three out of the 32 epochs have relatively low S/N (S/N $< 0.7 \langle S/N \rangle$) due to poor observing conditions and/or the inability to obtain the full 2-hr exposure.  \citet{Sun15b} omitted these epochs from their investigation, as they would have systematically biased the measured quasar structure function.  We include these low-S/N epochs, as our goal is to analyze spectra over a wide range of S/N.}  However, in a small number of cases, the spectra were too noisy to even detect emission lines, and so these were omitted from subsequent analysis.  This occurred most frequently for Epoch 7, the lowest S/N epoch in the campaign.  Our final sample consists of 482 sources.  Most (405) QSOs have all 32 epochs of spectra included in our analysis, and only 9 QSOs have more than 3 (10\%) epochs discarded. Figure~\ref{F_physprop} shows the redshift and $i_{\rm psf}$ magnitude distribution of our sample.

\section{Spectroscopic Analysis}
\label{S_SpecAnal}

The SDSS-RM project provides a unique opportunity to study data-quality biases in the measurement of spectroscopic properties of quasars from survey spectra. Each epoch is representative of a typical survey-quality spectrum.\footnote{Note, the exposure time of our SE spectra is greater than a typical SDSS or BOSS quasar spectrum, with a 2-hr rather than the minimum 45-min exposure.  However, these longer exposures are offset by our deeper flux limit and so lead to a sample of SE spectra with similar S/N distributions to SDSS or BOSS quasars, where the SDSS spectral S/N limits are set by the successive 15-min exposures being halted when the S/N per pixel exceeded 4 for fiber magnitudes of $g=20.2$ and $i=19.9$ and BOSS was similarly limited by exposures stopping when the (S/N)$^2$ exceeded 22 for $i=21$ and 10 for $g=22$.}  We thus analyze all individual epochs to measure a distribution of spectral properties expected for SE spectra of a single object.  Throughout this work, we use the median of these distributions for each object as our ``SE measurement" with the measurement uncertainty defined as half of the 16$-$84\% inter-percentile range (HIPR) of the distribution, which would correspond to 1-$\sigma$ if the distribution were Gaussian.  We then combine all the spectra to make a single, high-S/N, ``coadded" spectrum using the latest BOSS spectroscopic pipeline {\it idlspec2d} \citep[see][and Schlegel et al., in prep.]{Shen15b}.  

\begin{figure}
\epsscale{1.2}
\plotone{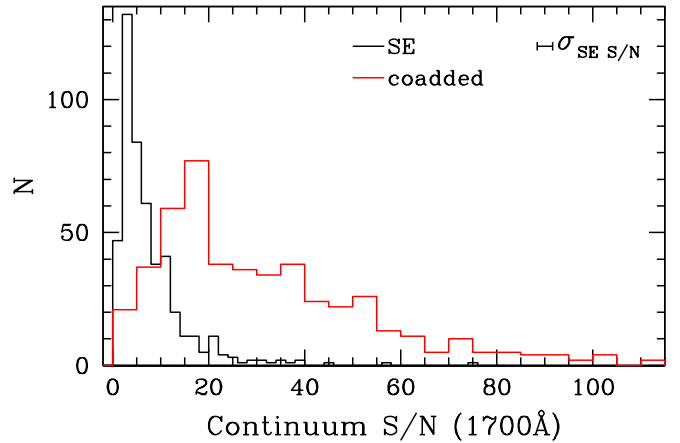}

\caption{Distribution of S/N of SE spectra (black) and coadded spectra (red) for our sample of 482 SDSS-RM quasars.  The coadded S/N is measured per Angstrom in an emission-line-free continuum window near restframe 1700\AA\ of each coadded spectrum, and the SE S/N shown is the median of the distribution of S/N measurements made similarly from all good SE spectra for each object.  The $\sigma_{\rm SE}$ error bar in the upper right corner represents the median of the distribution of scatter measurements made from the SE S/N distributions for all objects.  Five coadded spectra have S/N $>$ 115.} 

\label{F_snrprop}
\end{figure}

\begin{figure*}
\epsscale{1.}
\plotone{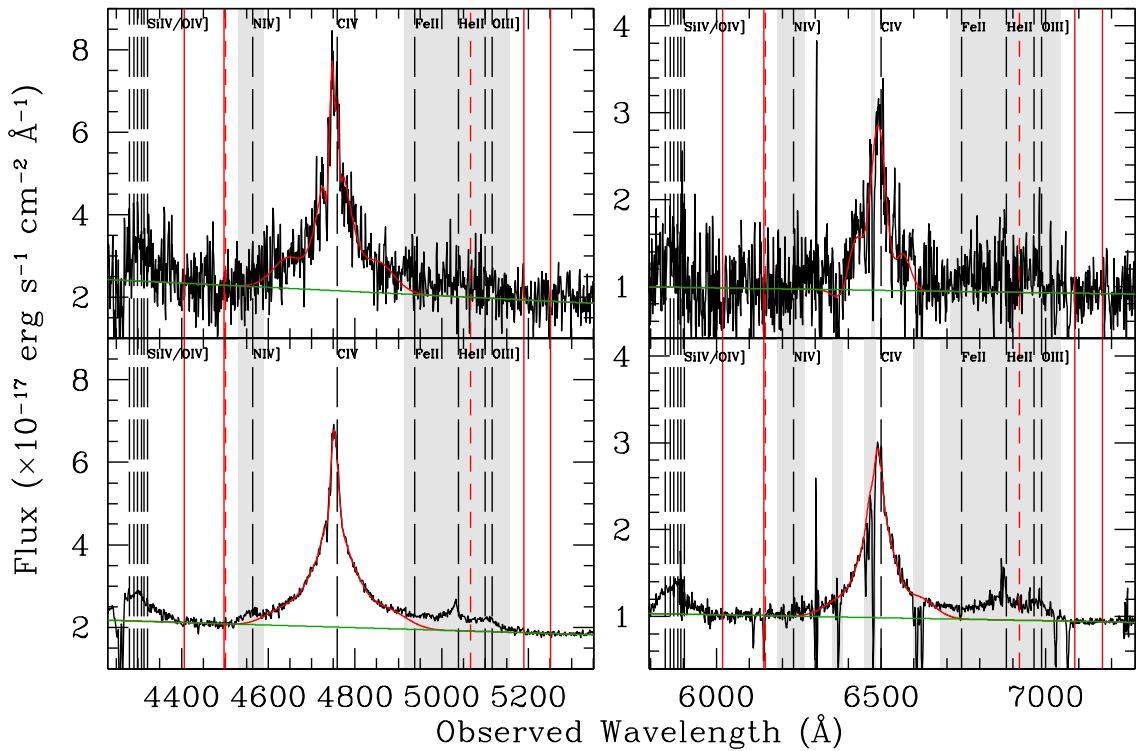}

\caption{Two examples (left: RM161 and right: RM139) of SDSS-RM SE (top panels) and coadded spectra (bottom panels) are shown in black. The best GH fits to the \civ\ emission lines are shown in red, and the best linear continuum fit is in green.  The vertical black dashed lines show the expected location of the labeled emission lines based on the SDSS-pipeline redshift for each object.  The red vertical solid lines show the boundaries of the two wavelength regions used to fit the continuum beneath the \civ\ emission, and the red vertical dashed lines bound the wavelengths over which the \civ\ line properties are measured. Shaded regions were masked during the fitting.} 

\label{F_exampspec}
\end{figure*}

We define the S/N per Angstrom, determined by the ratio of the mean flux to the dispersion in the flux in an emission-line-free continuum window near restframe 1700\AA. We use the same red continuum window as for the fits described below.  The coadded spectra have S/N 5$-$6 times higher than that of the SE spectra of each object.  Figure~\ref{F_snrprop} shows the S/N distributions of the SE and coadded spectra for our sample.  The median SE (coadded) spectrum continuum S/N is 5.5 (25.7) with a scatter, defined by the HIPR, of 4.9 (20.9).  Figure~\ref{F_exampspec} shows examples of the \civ\ emission line region in two SE spectra of SDSS-RM sources, with S/N near the sample median (top panels), and the corresponding coadded spectra (bottom panels). The coadded spectrum S/N is 48 (46) for the object in the left (right) panel. In both cases, the median S/N of all SE spectra of these objects is $\sim$8, but the SE spectra shown in Figure~\ref{F_exampspec} were chosen from an epoch with a S/N close to the median of the full sample (S/N$\sim$5).  

Biases in our coadded spectra due to intrinsic variability are unlikely, given the relatively high luminosity and high redshifts of our sample combined with the short campaign period. Our 6-month campaign duration covers only $\sim 1-3$ months in the rest frame of our $z>1.46$ quasars, and quasar (i) variability amplitudes and (ii) reverberation time-delays scale with luminosity inversely and directly, respectively.  Such variability, even if present on these timescales, would only change the overall line flux, to first order.  Gross profile change or velocity-dependent flux changes occur on timescales not probed by our campaign \citep[see, e.g.,][]{Sergeev07, Shen15b}.

\subsection{CIV line profile fits}

Because we are only investigating properties of the \civ\ emission line in this work, we only fit this localized region of the spectra.  Our general procedure for fitting the \civ\ emission line follows that described by \citet{Denney13}.  We first fit a linear continuum beneath the line, anchored by the mean flux in a blue and a red continuum region on either side of \civ\ (see Fig.~\ref{F_exampspec}).  While the AGN continuum is best described by a power law over longer wavelength ranges, there is very little difference locally (i.e., under a single emission-line region), between using a linear continuum and a power law.  We then mask the wavelength regions covering rest-frame 1475$-$1495\AA\ and 1600$-$1680\AA\ to exclude possible contributions from blended \niv\ $\lambda$1486 line emission and the ``red shelf" emission often observed between \civ\ and \heii\ $\lambda$1640 \citep[see][]{Fine10, Assef11, Denney13}.  We then fit the \civ\ emission line with 6th order Gauss-Hermite (GH) polynomials that adopt the functional forms of, e.g., \citet{Cappellari02}, and the normalization of \citet{vanderMarel93}.  Iterative sigma-clipping was employed in fitting both the continuum and the \civ\ emission line to exclude spurious flux contributions to the fit, typically from noise, poorly-subtracted skyline residuals, or narrow, unmasked absorption features.

We prefer the GH polynomial fits for \civ\ because they model the typically non-symmetric and non-Gaussian profile of \civ\ with fewer components than required when using (for example) only Gaussian functions.  However, it is possible that in low S/N data, the freedom of the GH polynomials may lead to fit profiles with more structure than is likely real, since a lower $\chi^2$ can be achieved by matching the fit to what are ultimately spurious emission signatures ascribable to noise. We do not ascribe physical meaning to any of the fit components.  We simply aim to reconstruct the best overall model of the intrinsic, unblended, unabsorbed profile.  We added fit components iteratively, only adding additional components when residual emission line flux from any part of the line --- usually the peak --- remained above the average continuum noise level after subtracting each fit component.  


Spectra were fit interactively, but with an automated pipeline to better reproduce typical literature practices \citep[e.g.,][]{YShen08, Shen11}. The automated pipeline set the continuum regions at restframe wavelengths 1435$-$1465\AA\ and 1690$-$1710\AA, although adjustments were made for the lowest redshift objects where the blue region fell at the noisy edge of the spectrum, for the presence of significant absorption from inspection of the first-epoch SE spectrum or the coadded spectrum, or for chip defects in any spectrum.  Next, the region over which \civ\ was fit was set to rest frame wavelengths 1466$-$1650\AA. This range was intentionally chosen to be broad enough to encompass all possible line widths.  The \civ\ line properties are measured directly from the best-fit profile, so the usual blended emission that may fall within these wavelengths, such as from \niv\ or \heii, is masked during fitting and does not contribute to the final fit.  

Manual intervention in the fitting procedure was necessary for several reasons.  First, fitting was interrupted to change or identify additional masking needs --- most often for absorption. The identification and details of contaminating features are more easily discerned in higher S/N spectra, so a more hands-on approach to the fitting was allowed when the additional signal provided this information.  As such, modifications to the boundaries and masks were optimized on an object-to-object basis.  This was often possible only for the coadded spectra, but when applied to the SE spectra, any masking or boundary changes were set by only the first-epoch spectrum and then kept the same for all additional epochs. Common modifications were to widen the red shelf mask further to the blue or the size and/or location of the \niv\ and continuum masks to more optimal spectral regions.  More detailed absorption line masks were also often needed in higher S/N spectra because weak absorption features were not always cleanly removed by the sigma-clipping procedure. Manual intervention was also used to confirm when multiple components were needed to fit the profile. This was done for both the coadded spectrum and for every SE spectrum.  Finally, the interactive process also allowed us to flag and then remove previously unidentified ``dropped" spectra from our analysis.  These interactive steps likely increase the robustness of our fits compared to fully automated analyses of large survey samples for which it is infeasible to visually inspect the fits to every spectrum.   

To prevent any unconscious bias in the way the SE spectra were fit or masked for contaminating features, we fit the SE spectra before fitting the coadded spectra.  To do otherwise might bias our analysis because using the information from the higher S/N coadded spectra to inform treatment of the lower S/N SE spectra is a luxury not afforded in typical analyses of survey-quality data. Examples of GH fits are shown with the spectra in Figure~\ref{F_exampspec}.

\subsection{Measurements of \civ\ Emission Line Properties}

We focus our attention on only the two most commonly employed emission-line width characterizations used to derive BH masses: the FWHM and the line dispersion ($\sigma_l$), the square root of the second moment of the line profile.  However, we also consider the velocity width characterized by the mean absolute deviation (MAD\footnote{We note that our definition of the MAD is not the same as the standard definition of the MAD as the {\it median} absolute deviation from the distribution median.}) from the median velocity, defined following the notation of \citet{Peterson04} as

 \begin{equation}
 {\rm MAD} = \int | \lambda - \lambda_{\rm med} | P(\lambda) d\lambda\  \big/
 \int P(\lambda) d \lambda,
 \end{equation} 

\noindent where $P(\lambda)$ is the continuum-subtracted emission-line profile, and $\lambda_{\rm med}$ is the flux-weighted median of the profile.

Additional properties of the \civ\ emission line we investigate include (i) its peak wavelength, defined from the centroid of the pixels $\geq$\,95\% of the peak line flux, (ii) the equivalent width (EW), defined here with respect to the monochromatic continuum flux level measured in the continuum window covering rest frame $\sim$1450\AA, and the line ``shape", characterized by (iii) the line kurtosis and (iv) the ratio FWHM/$\sigma_l$.  The amount by which the peak of \civ\ is blueshifted with respect to the systemic is often an additional line property of interest, but the determination of the blueshift depends on the reliability of the systemic redshift.  Denney et al. (2016, in preparation) investigates the potential for biases in the redshifts determined for this sample, and we therefore defer discussions of the \civ\ blueshift to that work. 

Line properties are measured from the continuum-subtracted best-fit GH polynomial fits.  Any pixel with negative flux after the continuum subtraction is not included in the calculation of any line property.\footnote{These are zero-noise profile fits. Negative pixels are therefore not due to noise characteristics, in which case they would be approximately balanced by positive pixels and their values should remain in the calculation.  Instead, the fit sometimes produces negative pixels in masked regions (e.g., over absorption or the \civ\ redshelf) where the profile is unconstrained.  Alternatively, it could occur in cases where absorption was not accurately masked due to noise (see, e.g., the top right panel of Figure~\ref{F_exampspec}).} Individual spectra were not used when the measurement of any of these quantities were bogus, e.g., when an undefined EW measurement was returned because the mean continuum flux in the 1450\AA\ region was found to be $\leq 0$, even if the fit found a weak line with a defined width, or when identification of the \civ\ emission line failed completely due to low S/N, resulting in a null line width and other properties.  After dropping all such cases, the subsequent analysis is based on the 482 coadded spectra and 15275 SE spectra from all ``good epochs".

\begin{figure}
\epsscale{1.2}
\plotone{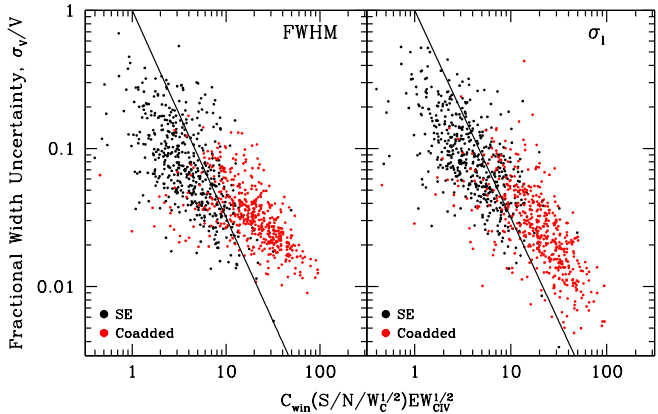}

\caption{Fractional FWHM (left) and $\sigma_l$ (right) uncertainties for the SE (black points) and coadded spectra (red points) as a function of the signal-to-noise ratio of the line, estimated as C$_{\rm win}$S/N$\sqrt{EW/W_C}$, where $W_C$ is width of the continuum window over which the S/N was measured, and C$_{\rm win}$ is a normalization factor related to the sizes of the continuum and line integration windows and the EW of the line.  The S/N is measured per Angstrom in an emission-line-free continuum window near restframe 1700\AA.  The solid black line is the expected reference relation, $\sigma_V/V \propto$(S/N$\sqrt{EW/W_C})^{-1}$.} 

\label{F_widtherrors}
\end{figure}

Similarly to the SE spectra S/N measurements, the SE line property ``measurement" for each quasar is taken to be the median of the distribution of all the measurements from all good epochs of that source, and the measurement uncertainty is taken as the HIPR of the distribution.  While all spectra nominally represent equivalent and independent observations of each source, we use the median and HIPR as opposed to the mean and standard deviation of these distributions to help account for possible outliers due to variable observing conditions between epochs.  The emission line property measurements for all objects based on the coadded spectra are made directly from the profile fit to the coadded spectrum for that source.  The uncertainties in these measurements are estimated by performing Monte Carlo simulations using the fit to the coadded spectrum.  We create 500 mock spectra by resampling the flux in the GH polynomial fit based on Gaussian deviates derived from the coadded error spectrum for each source.  We assign the uncertainty in the measured line properties of each coadded spectrum to be the standard deviation about the mean of the distribution of 500 measurements of that property made from the mock spectra.   Figure~\ref{F_widtherrors} shows the relative line width uncertainties, $\sigma_V/V$, for the FWHM and $\sigma_l$ of the SE and coadded spectra with respect to an approximated line S/N, for which we expect a linear relation.  We find that line width uncertainties can be a significant fraction of the measured line width in the lowest S/N spectra, and there is a factor of 2$-$3 increase in the magnitude of the statistical line width measurement uncertainties between high- (S/N$>$10) and low- (S/N$<$5) quality data for all characterizations of the \civ\ velocity width.  On the other hand, while the relative uncertainty drops to $\lesssim 5\%$ for spectra with S/N $\gtrsim$10$-$20, we find the trend is shallower than the expected slope of -1 following $\sigma_V/V \propto$(S/N)$^{-1}$(EW/$W_C$)$^{-1/2}$, given that the line photons dominate the continuum photons and where $W_C$ is the width of the continuum window over which the S/N was measured.  This is likely because of the details of the FWHM and $\sigma_l$ calculations. The FWHM does not depend on the whole line, but rather only on a fraction of the pixels (i.e., those near the peak and the 50\% flux level).  The shallower trend with $\sigma_l$ is likely because $\sigma_l$ is relatively more weighted by the flux in the wings of the line, where the line flux is lower.  Table \ref{T_widths} lists the uncertainty distribution properties for the full sample and various sub-samples divided by S/N cuts.      

\section{Discussion}
\label{S_Discussion}

We compare the line-property measurements between the ``highest-S/N" (coadded) spectra and the ``lower-S/N" (SE) spectra by looking at the measurement differences, $\Delta$(X) = X(coadded)$-$Median[X(SE)], in the line property, X, as a function of data quality.  The S/N of a small percentage of the coadded spectra are also ``low", compared to the median of the coadded spectrum S/N distribution (see Fig.\ \ref{F_snrprop}), and fall within the lowest quality subsample we consider in our analysis (S/N$<$5).  The coadded spectrum is, nonetheless, the highest-S/N observation of a given source and therefore always contains more information than a SE spectrum. 

\subsection{Data-Quality Biases in Common Velocity-Width Characterizations}
\label{S_DiscEmLineBias}

\begin{figure*}
\epsscale{0.9}
\plotone{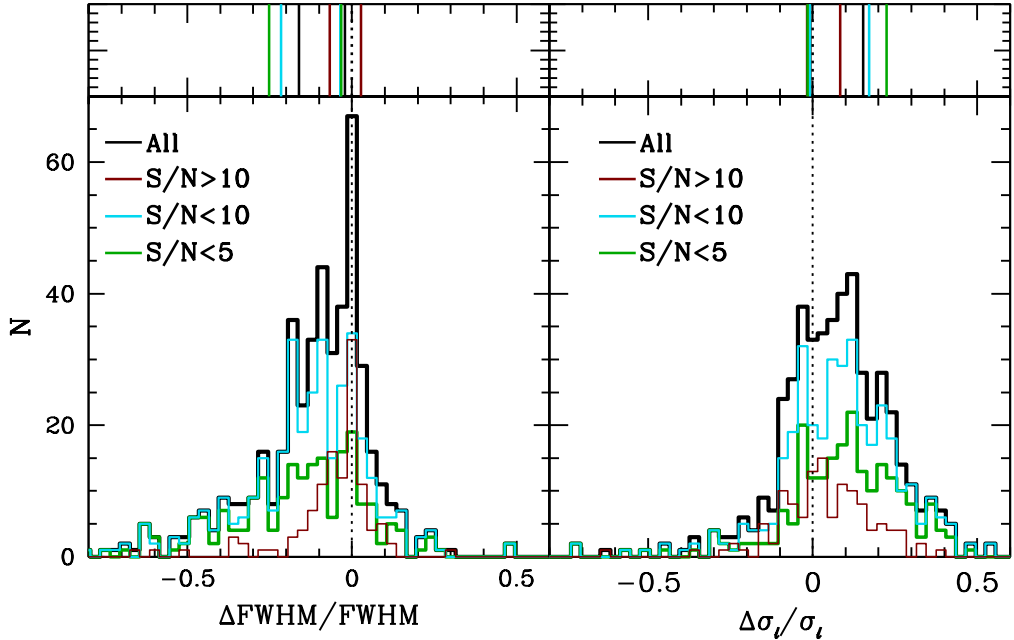}

\caption{Histograms of the fractional differences between coadded and SE \civ\ emission-line width measurements, FWHM (left) and $\sigma_l$ (right).  The histograms are color-coded by data quality as shown in the legend (note: the S/N$<$5 sub-sample is contained within the S/N$<$10 subsample). The similarly color-coded pairs of lines in the small top panels are centered on the median of each data quality distribution, given in Table \ref{T_widthdiffs}, and the range covered corresponds to the distribution median $\pm$ the median fractional uncertainty given in Column 5 of Table \ref{T_widths}. An increasing bias is seen for lower S/N subsamples that is encompassed within the statistical uncertainties at all S/N levels for $\sigma_l$ but only for S/N$>$10 for the FWHM.} 

\label{F_widthscomparehist}
\end{figure*}

\begin{figure*}
\epsscale{0.9}
\plotone{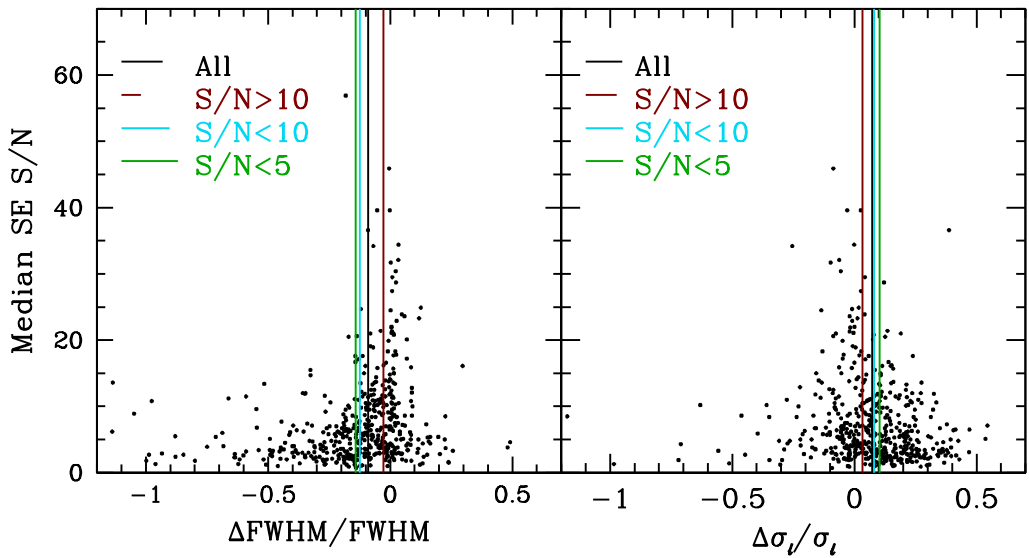}

\caption{Distribution of data-quality-related biases in the line width measurements. The left (right) panel shows the SE S/N as a function of the fractional difference between the coadded and SE FWHM (line dispersion, $\sigma_l$) of the \civ\ emission line. The vertical lines show the median velocity width difference of each histogram with the same color shown in Figure~\ref{F_widthscomparehist} and shown in the legend, and length of the lines within the legend represents the HIPR range of each respective distribution.} 

\label{F_widthscompareSNscatt}
\end{figure*}

We first investigate how data quality affects the distributions of the fractional line width differences, $\Delta$(X)/X(coadded), for both the FWHM and line dispersion, $\sigma_l$.  The left panel(s) in Figures \ref{F_widthscomparehist} and \ref{F_widthscompareSNscatt} show the results for the fractional FWHM differences, $\Delta$FWHM/FWHM. Table \ref{T_widthdiffs} gives the statistics of these distributions.  We find a systematic bias such that the FWHM is overestimated in the SE spectra as compared to the coadded spectra.  The magnitude of this bias, quantified by the median of the $\Delta$FWHM/FWHM distributions, is within the line width uncertainties only for the SE S/N$>$10 subsample, similar to other studies \citep{Denney09a, Shen11}.  While the bias is only marginally larger than the line-width uncertainties for the full sample and both subsamples with SE S/N$<$10, the distribution HIPR width, i.e., the scatter, increases by a factor of two from high to low S/N and has a significant asymmetry that suggests very large biases are possible, much larger than the formal measurement uncertainties, for some FWHM measurements.  Inspection of the left panel of Figure~\ref{F_widthscompareSNscatt} demonstrates that this asymmetry and the magnitude of the systematic bias are anticorrelated with spectral S/N.  

The right panel(s) in Figures~\ref{F_widthscomparehist} and \ref{F_widthscompareSNscatt} and the statistics in Table \ref{T_widthdiffs} show the results for the line dispersion, $\Delta\sigma_l$/$\sigma_l$.  A small systematic bias also exists in the line dispersion measurements, but in the opposite sense --- the line dispersion tends to be underestimated in lower S/N spectra.  The magnitude of this bias also increases with decreasing S/N. However, unlike the systematic bias in the FWHM measurements, the measurement uncertainties encompass the observed systematic shift in the SE line dispersion measurements at all spectral S/N levels.  The $\Delta\sigma_l$ distributions are also more symmetric than the $\Delta$FWHM distributions, though the right panel of Figure~\ref{F_widthscompareSNscatt} and statistics in Table \ref{T_widthdiffs} demonstrate that data quality still contributes to both the systematic bias and the broadening of the $\Delta\sigma_l$ distributions at lower S/N.  We caution the reader, however, that these results hold only when line dispersion measurements are made self-consistently across a sample.   Larger DC offsets in line dispersion measurements, and thus BH mass estimates, can arise {\it between} samples that are analyzed with different spectral processing methods, i.e., continuum and line boundary placements and emission deblending assumptions \citep[e.g.,][]{Denney09a, Fine10, Assef11}.  Such offsets can be surmounted by following the same spectral processing method used in the calibration of the desired mass scaling relationship \citep{Vestergaard06, Park13}. 

\subsection{Data-Quality Biases in Other Line Properties}

\begin{figure*}
\epsscale{0.9}
\plotone{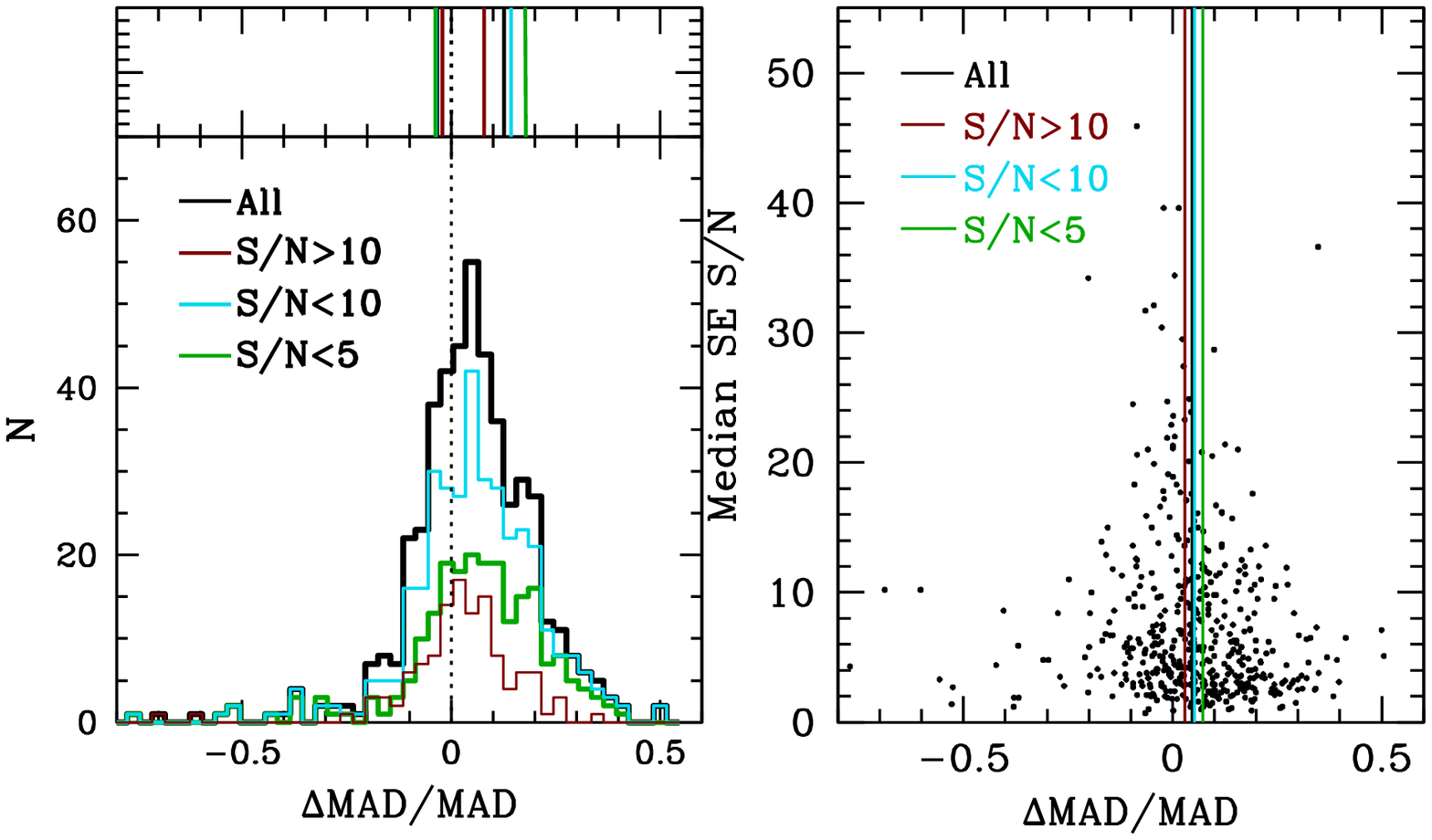}

\caption{{\it Left}: Same as Figure \ref{F_widthscomparehist} but for the MAD.  {\it Right}: Same as Figure \ref{F_widthscompareSNscatt} but for the MAD.} 

\label{F_plotMADdistr}
\end{figure*}

We also investigate the data-quality dependence of the other line properties we measured above --- the \civ\ line MAD, EW, shape (kurtosis and width ratio), and centroid (see Tables \ref{T_widths} and \ref{T_widthdiffs}).  The left panel of Figure~\ref{F_plotMADdistr} shows that the MAD behaves most similarly to $\sigma_l$ in that it increasingly underestimates the width with decreasing S/N, although the relative magnitude of the bias is smaller than that observed for both FWHM and $\sigma_l$, and the systematic offset stays well within the measurement uncertainties for all S/N subsamples.  The MAD number distributions as a function of $\Delta$MAD (right panel of Fig.~\ref{F_plotMADdistr}) are also relatively more symmetric with less extended wings in either direction compared to the FWHM and $\sigma_l$. This suggests it may be a more robust characterization of the velocity width in low-quality data.  

The trends with decreasing S/N differ between the other three properties we investigated.  The \civ\ line EW is systematically underestimated in the low S/N spectra, but the degree of systematic bias is within the larger measurement uncertainties, which is consistent with the findings of \citet{Shen11}.  For the \civ\ centroid, a factor of $\sim$2 increase in statistical uncertainty is seen between the highest and lowest S/N subsamples, but a significant systematic bias is not detected for any S/N subsample, demonstrating it to be the most robust of all properties we investigated in this respect (see Shen et al.\ 2016, in prep., for similar results using this sample).  The kurtosis of the \civ\ line, however, is the least robust, which is not altogether surprising given its construction from relatively higher moments of the line profile. The formal measurement uncertainty only increases marginally with decreasing S/N, but there is a statistically significant, $>$3$\sigma$, systematic bias in $\Delta$Kurtosis for all S/N subsamples we investigate. This bias improves somewhat when we characterize the \civ\ profile by the FWHM/$\sigma_l$ ratio.  The measurement uncertainties increase by the same amount as the line width uncertainties --- a factor of $\sim$2$-$3 between the S/N$>$10 and S/N$<$5 subsamples --- as expected by construction, and the shape is systematically overestimated in lower S/N spectra.  This bias is within the 1-$\sigma$ measurement uncertainties of only the S/N$>$10 subsample but still within $\sim$1.5$\sigma$ for lower S/N, demonstrating it to be a more robust way to characterize the \civ\ profile than the kurtosis.

\begin{figure*}
\epsscale{1.}
\plotone{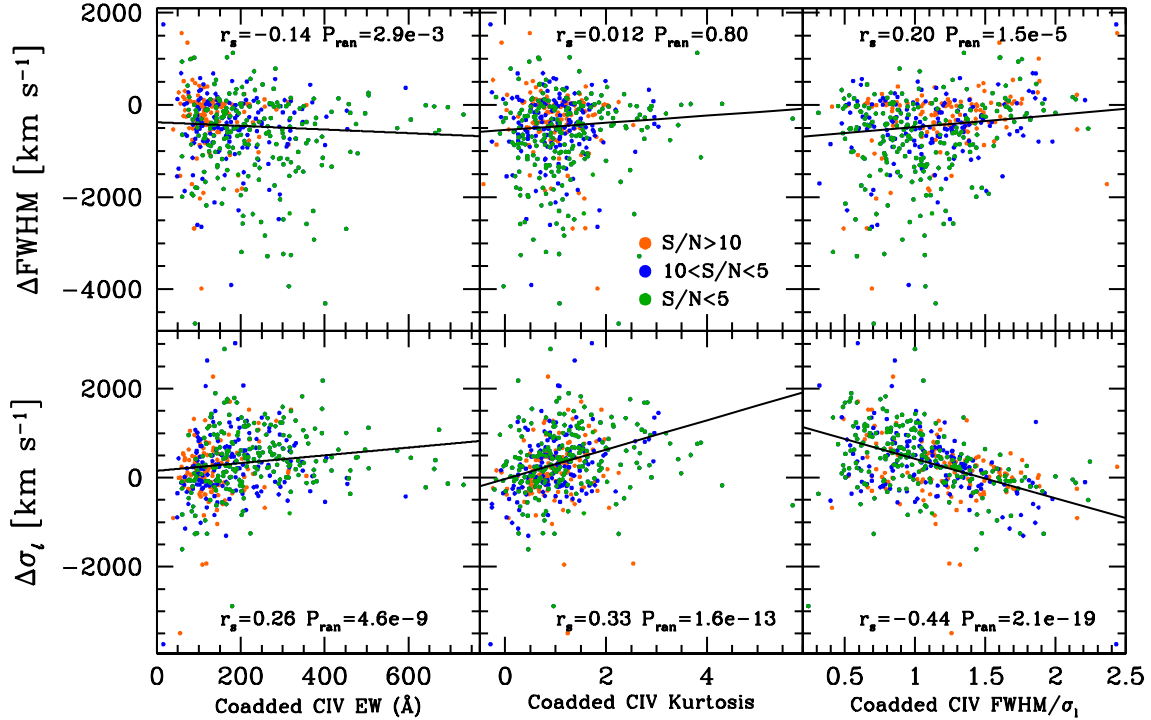}

\caption{\civ\ line width differences as a function of other \civ\ emission line properties.  The top (bottom) panels show results based on the FWHM (line dispersion, $\sigma_l$).  The left, middle, and right panels show the dependence of the difference on the \civ\ EW, kurtosis, and shape (FWHM/$\sigma_l$), respectively.  The points are color-coded by the varying data quality of each subsample:  S/N$>$10 (orange), 10$<$S/N$<$5 (blue), and S/N$<$5 (green). The solid black lines show the best fit linear regression fit. The Spearman rank correlation coefficient, $r_s$, and the probability that a correlation is found by chance, $P_{\rm ran}$, are given in each panel.} 

\label{F_linepropquality}
\end{figure*}

We also investigate the possible dependence of the line-width differences on other line properties.  Figure~\ref{F_linepropquality} shows the dependence of $\Delta$FWHM (top panels) and $\Delta\sigma_l$ (bottom panels) on the \civ\ EW (left), kurtosis (middle), and the FWHM/$\sigma_l$ ratio (right). We performed a formal linear-regression analysis on the dependencies using LINMIX\_ERR \citep{Kelly07b} and quantify the results with a Spearman rank order test.  We evaluate the significance of any correlations using the Spearman rank correlation coefficient, $r_s$, the probability that a correlation is found by chance, $P_{\rm ran}$, and the formal uncertainties on the slope of the regression fit.  

The only statistically significant trend with $\Delta$FWHM is with the FWHM/$\sigma_l$ ratio, but the correlation slope is consistent with a slope of zero within the 3$\sigma$ level.  The correlation appears primarily driven by the large scatter seen only at small values of the FWHM/$\sigma_l$ ratio. Direct inspection of the data (see, e.g., Figure~\ref{F_exampspec}) suggests that the main driver for the systematic overestimation of the FWHM in lower S/N data is an underestimation of the emission line peak due to noise.  This is also consistent with the findings of \citet{Denney09a} for \Hbeta\ FWHM measurements.  With \civ, however, additional systematics contribute at low S/N in cases where absorption cannot be accurately characterized.  The peak and/or half-maximum flux width can thus be under- or over-estimated, on a case-dependent basis, leading to additional dispersion in the $\Delta$FWHM distribution that may obscure other more subtle trends.  

\begin{figure}
\epsscale{1.}
\plotone{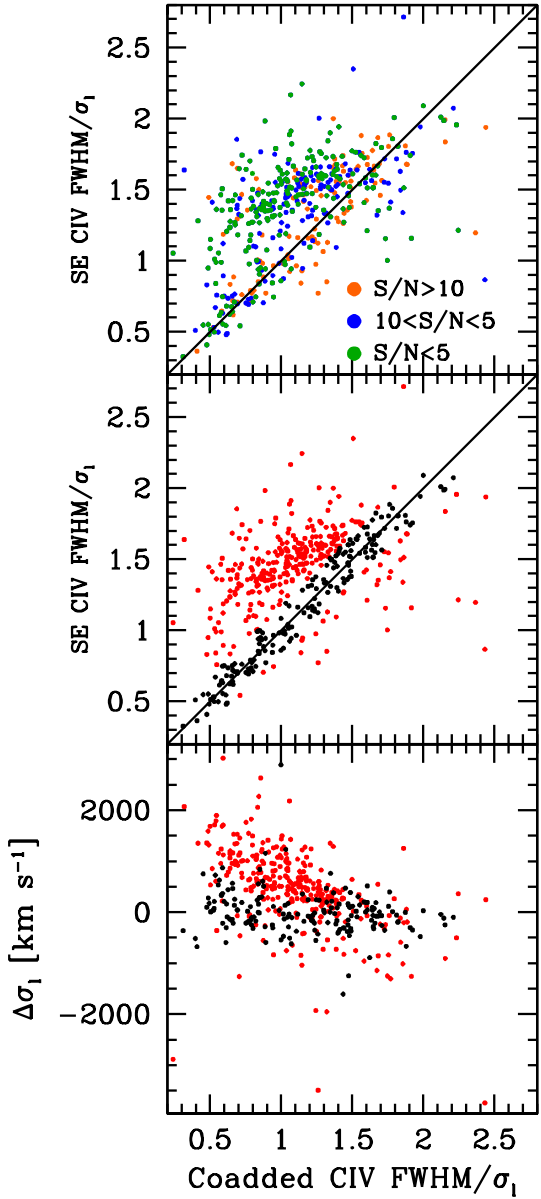}

\caption{Trends between the \civ\ shape parameter (FWHM/$\sigma_l$) and the coadd$-$SE line dispersion differences. The top panel compares the SE \civ\ line shape measurements to those measured from the coadded spectra; colors represent different S/N subsamples and are the same as in Figure~\ref{F_linepropquality}.  The middle panel is the same as the top panel, except the color coding now reflects the division of objects into two populations:  the black (red) points represent objects for which the SE shape is (is not) roughly consistent with the coadded shape.  The black points are defined by objects within the 1$\sigma$ scatter of having the same shape (black solid line), where 1$\sigma$ is defined from only points below this relation.  All other objects are shown in red.  The bottom panel is the same as the bottom right panel of Figure~\ref{F_linepropquality} but uses the same color coding as the middle panel.} 

\label{F_coaddSEshapecompare}
\end{figure}

The trends in Figure~\ref{F_linepropquality} between $\Delta\sigma_l$ and the other \civ\ line properties are all statistically significant and have regression fit slopes that deviate from zero by more than 3$\sigma$.  There does not appear to be any strong dependence on data quality --- all data quality subsamples follow the same trends.  We focus additional attention on the correlation with the FWHM/$\sigma_l$ ratio because is has the highest significance and the smallest scatter.  We first investigate possible systematic differences between shape measured from the coadded versus SE spectrum.  The top panel of Figure~\ref{F_coaddSEshapecompare} suggests objects can be roughly divided into (i) objects along the line of equal SE and coadded \civ\ shape, regardless of S/N, and (ii) a cloud of points with significantly different SE and coadded shape measurements, although there are points scattered between the two groups for low values of the coadded shape, and the two groups merge for large shape values.  The latter population is dominated by the lowest S/N subsample, but still includes some objects with reasonably high S/N.  

We separate these two populations in the middle panel of Figure~\ref{F_coaddSEshapecompare} by color-coding them by their deviation from having the same shape measurements. The bottom panel of Figure~\ref{F_coaddSEshapecompare} shows this same population division in the $\Delta\sigma_l$--coadded \civ\ shape parameter space (same as the bottom right panel of Figure~\ref{F_linepropquality}).  This division demonstrates that the observed trend is due to objects for which the \civ\ profile is not accurately characterizable once the S/N is degraded, even marginally, as we see deviations of objects from all S/N subsamples, but this is a problem for only some quasars, which leads to the relatively larger scatter for smaller FWHM/$\sigma_l$ ratios.

Inspection of individual cases suggests that this bias (and therefore the population separation with shape) occurs when the wings of the emission line become mischaracterized because of contamination from the noise of the continuum.  The predominant effect is that the profile defined by the GH polynomial fits becomes truncated at artificially lower velocities than inferred from the higher-S/N coadded spectrum. It is also possible that the continuum level is more accurately characterized in higher S/N data, leading to lower overall accuracy of the profile fit in the SE spectrum as compared to the coadded.  This former effect can clearly be seen in the example spectra shown in the left panels of Figure~\ref{F_exampspec}.  Another possible contributor to artificially truncated profile fits is blending of the \civ\ wings with the red shelf of \civ.  The profiles most affected by premature truncation due to both noise and blending are those with small values of the shape parameters, which corresponds to the profiles that are very `peaky' with strong narrow velocity cores and very broad wings.  This bias is likely the driver for the trends between $\Delta\sigma_l$ and the other line properties shown in Figure~\ref{F_linepropquality}, as well.

\section{Summary}
\label{S_Conclusions}

We have investigated data-quality related statistical uncertainties and systematic biases in spectroscopic properties measured for high-redshift quasars.  We have been particularly interested in properties of the \civ\ emission line, as this line is of interest for high-redshift BH-mass estimates.  This investigation was only possible because of the availability of spectroscopic monitoring data taken by the SDSS-RM Project.  Taken individually, these spectra are representative of typical survey-quality data; however, coadding produces a higher S/N spectrum, which can be used to make direct comparisons of the measurements of spectral properties between low and high-S/N spectra.

Our analysis shows that the statistical measurement uncertainties of all the \civ\ velocity width characterizations we consider (FWHM, $\sigma_l$, and MAD) increase significantly, by a factor of 2$-$3, when going from high to low quality data (see Table \ref{T_widths}, Columns 4 and 5, and Fig.~\ref{F_widtherrors}).  The statistical uncertainty in the BH-mass estimates, which depend on the line width squared, are therefore a factor of two larger than this.  We also find the following systematic trends for the line-width characterizations we consider:

\paragraph {\bf FWHM} At low S/N, a systematic overestimation of the \civ\ FWHM measurements is introduced, although it is not significantly larger than the measurement uncertainties, on average. The observed bias can be attributed to inaccuracies in characterizing the emission-line peak flux level in noisy data (see Fig.\ \ref{F_widthscompareSNscatt} and Table 2).  Uncharacterized absorption can exacerbate the FWHM biases \citep[see also,][]{Assef11, Denney13}, and the systematic bias can be severe in some cases.  Interestingly, similar S/N investigations into potential biases in FWHM measurements described by \citet{Shen11} for a different sample suggest that the FWHM in low S/N data may be systematically underestimated when the profiles are described by multi-Gaussian fits, so details of the bias may depend on the specific functional form used to fit the data. 

\paragraph {\bf Line Dispersion} Biases in $\sigma_l$ at low S/N are less than those in the FWHM and scale with the statistical uncertainty.  When present, underestimation of $\sigma_l$ is caused mainly by the inability to accurately fit the emission line wings in the presence of a noisy continuum.   As a result \civ\ profiles with `peaky' cores and extended wings are likely to be biased more significantly than `stumpy' or `boxy' profiles without extended wings (Figs.\ \ref{F_linepropquality} and \ref{F_coaddSEshapecompare}).

\paragraph {\bf MAD} The systematic bias in the MAD measurements was relatively smaller than for the FWHM and $\sigma_l$. The bias was also always well within the statistical uncertainties, and the distribution of $\Delta$MAD measurements remains relatively symmetric and centrally peaked with decreasing S/N (see Fig.\ \ref{F_plotMADdistr}).  
\vspace{0.2cm}

These trends of FWHM and line dispersion measurements with S/N are consistent with the results for \Hbeta\ from similar analyses presented by \citet{Denney09a}, but that study did not investigate the MAD.  We conclude here that the MAD is the most reliable measure of the velocity width for low-quality data.  Nonetheless, further analysis is needed to investigate how good of a proxy this characterization is for the virial BLR velocity \citep[see][]{Peterson04}, and SE BH mass scaling relationships have not yet been developed and calibrated for this characterization. 

We also stress to the reader that the present study has only been focused on biases due to data-quality considerations.  We make no preference for which line-width characterization is a {\it better} proxy for the reverberating BLR velocity dispersion that traces the gravitational potential of the BH.  The FWHM is often preferred because it is more simple to measure and less susceptible to the subjectiveness of deblending procedures.  On the other hand, recent studies \citep{Assef11, Denney12, Denney13} have demonstrated that $\sigma_l$ is less biased than FWHM for estimating \civ\ BH masses, which is least partially attributable to the presence of non-variable flux contributions to the \civ\ emission line and/or a continuum color term that are yet unaccounted for in \civ-based BH mass scaling relation calibrations (see also \citealt{Rafiee&Hall11}, for similar \mgii\ trends, and \citealt{Peterson14}, for additional discussions).

\acknowledgements We are grateful for the editorial contributions to this work from C.~S.~Kochanek. KDD is supported by an NSF AAPF fellowship awarded under NSF grant AST-1302093. KH acknowledges support from STFC grant ST/M001296/1. WNB acknowledges support from NSF grant AST-1516784. LCH thanks Carnegie Observatories for providing telescope access and acknowledges financial support from Peking University, the Kavli Foundation, the Chinese Academy of Science through grant No. XDB09030102 (Emergence of Cosmological Structures) from the Strategic Priority Research Program, and from the National Natural Science Foundation of China through grant No. 11473002. BPM is grateful for support from NSF grant AST-10008882. Funding for SDSS-III has been provided by the Alfred P. Sloan Foundation, the Participating Institutions, the National Science Foundation, and the U.S. Department of Energy Office of Science. The SDSS-III web site is http://www.sdss3.org/. SDSS-III is managed by the Astrophysical Research Consortium for the Participating Institutions of the SDSS-III Collaboration including the University of Arizona, the Brazilian Participation Group, Brookhaven National Laboratory, University of Cambridge, Carnegie Mellon University, University of Florida, the French Participation Group, the German Participation Group, Harvard University, the Instituto de Astrofisica de Canarias, the Michigan State/Notre Dame/JINA Participation Group, Johns Hopkins University, Lawrence Berkeley National Laboratory, Max Planck Institute for Astrophysics, Max Planck Institute for Extraterrestrial Physics, New Mexico State University, New York University, Ohio State University, Pennsylvania State University, University of Portsmouth, Princeton University, the Spanish Participation Group, University of Tokyo, University of Utah, Vanderbilt University, University of Virginia, University of Washington, and Yale University.



\clearpage

\begin{deluxetable}{lccccccc}
\tablecolumns{8}
\tablewidth{7.0in}
\tablecaption{Data-quality Dependence of Distributions of SE \civ\ Line Property Statistical Uncertainties}
\tabletypesize{\scriptsize}
\tablehead{
\colhead{Distribution} &  \colhead{Sub-sample} & \colhead{Number} & \colhead{Median Sample\tablenotemark{a}} & \colhead{Uncertainty} & \colhead{Uncertainty} & \colhead{Uncertainty} & \colhead{Uncertainty}\\
\colhead{Property}&\colhead{Description} & \colhead{of Obj.} & \colhead{S/N} & \colhead{Median\tablenotemark{b}} & \colhead{HIPR\tablenotemark{b}}&\colhead{Mean\tablenotemark{b}} &\colhead{Std.\ Dev.\tablenotemark{b}}\\
\colhead{(1)} &
\colhead{(2)} &
\colhead{(3)} &
\colhead{(4)} &
\colhead{(5)} &
\colhead{(6)} &
\colhead{(7)} &
\colhead{(8)} 
}

\startdata
FWHM				& S/N$>$10	& 122	& 13.9	& 186 (4\%)	& 156	& 276 (5\%) 	& 259	\\
FWHM 				& All			& 482	& 5.5		& 340 (7\%) 	& 310	& 454 (10\%) 	& 403	\\
FWHM				& S/N$<$10	& 360	& 4.1		& 396 (9\%)	& 328	& 514 (11\%) 	& 425	\\
\vspace{0.2cm}
FWHM				& S/N$<$5	& 218	& 2.9		& 498 (11\%)	& 367	& 602 (13\%) 	& 478	\\
Line dispersion, $\sigma_l$ & S/N$>$10	& 122	& 13.9	& 185 (5\%)	& 128	& 247 (6\%) 	& 223	\\	
Line dispersion, $\sigma_l$ & All		& 482	& 5.5		& 285 (8\%)	& 227	& 379 (10\%) 	& 313	\\ 
Line dispersion, $\sigma_l$ & S/N$<$10	& 360	& 4.1		& 321 (9\%)	& 255 	& 424 (12\%) 	& 326	\\
\vspace{0.2cm}
Line dispersion, $\sigma_l$ & S/N$<$5	& 218	& 2.9		& 368 (12\%)	& 298	& 481 (15\%) 	& 336	\\
MAD					& S/N$>$10	& 122	& 13.9	& 133 (5\%)	& 88  	& 176 (6\%) 	& 181  		\\
MAD					& All			& 482	& 5.5		& 201 (8\%)	& 175 	& 277 (10\%)  	& 251  	\\
MAD					& S/N$<$10	& 360	& 4.1		& 226 (9\%)	& 187  	& 311 (12\%) 	& 262  \\
\vspace{0.2cm}
MAD					& S/N$<$5	& 218	& 2.9		& 259  (11\%)	& 212  	& 353 (14\%)	 & 267  \\
EW					& S/N$>$10	& 122	& 13.9	& 7 			& 5  		& 9  			& 7  		\\ 
EW					& All			& 482	& 5.5		& 17 			& 16 		& 27  		& 39  	\\
EW					& S/N$<$10	& 360	& 4.1		& 22 			& 18  	& 33  		& 43  \\
\vspace{0.2cm}
EW					& S/N$<$5	& 218	& 2.9		& 31  		& 22  	& 43  		& 52  \\
Centroid				& S/N$>$10	& 122	& 13.9	& 1.1 		& 1.0  	& 1.7  		& 2.1  \\ 
Centroid				& All			& 482	& 5.5		& 1.5  		& 1.4  	& 2.2  		& 2.2  \\
Centroid				& S/N$<$10	& 360	& 4.1		& 1.8 		& 1.4  	& 2.4  		& 2.3  \\
\vspace{0.2cm}
Centroid				& S/N$<$5	& 218	& 2.9		&  2.0 		& 1.6  	& 2.3  		& 2.5  \\
Kurtosis				& S/N$>$10	& 122	& 13.9	& 0.12  		& 0.08  	& 0.14  		& 0.09  \\ 
Kurtosis				& All			& 482	& 5.5		& 0.17  		& 0.12 	& 0.20  		& 0.14  \\
Kurtosis				& S/N$<$10	& 360	& 4.1		& 0.19  		& 0.14  	& 0.22  		& 0.15  \\
\vspace{0.2cm}
Kurtosis				& S/N$<$5	& 218	& 2.9		& 0.22  		& 0.15  	& 0.25  		& 0.16  \\
Shape (FWHM/	$\sigma_l$) & S/N$>$10	& 122	& 13.9	& 0.09  		& 0.05  	& 0.11 		& 0.09  \\ 
Shape (FWHM/	$\sigma_l$) & All		& 482	& 5.5		& 0.15  		& 0.13  	& 0.20  		& 0.17  \\
Shape (FWHM/	$\sigma_l$) & S/N$<$10	& 360	& 4.1		& 0.18  		& 0.14  	& 0.24  		& 0.18  \\
Shape (FWHM/	$\sigma_l$) & S/N$<$5	& 218	& 2.9		& 0.22  		& 0.17  	& 0.28  		& 0.20
\enddata

\tablenotetext{a}{The Median S/N is based on the distributions of SE spectra.  The S/N is measured per Angstrom, integrated over an emission-line-free continuum window, $\Delta W$, covering many resolution elements near restframe 1700\AA.}
\tablenotetext{b}{The median, HIPR, mean, and standard deviation (Std.\ Dev.) values are in units of \kms\ for the FWHM and $\sigma_l$ distributions and \AA\ for the EW and Centroid distributions.  The Kurtosis and Shape parameters are dimensionless. We also report the corresponding mean and medians of the fractional velocity distributions for SE Line Width Uncertainties in parentheses following each median and mean to be consistent with what is shown in Figures \ref{F_widtherrors}.}

\label{T_widths}
\end{deluxetable}

\begin{deluxetable}{cccccccc}
\tablecolumns{8}
\tablewidth{7.0in}
\tablecaption{Data-quality Dependence of Distributions of \civ\ Line Property Biases, $\Delta$X = X(Coadded)$-$X(SE)}
\tabletypesize{\scriptsize}
\tablehead{
\colhead{$\Delta$X Distribution} &  \colhead{Sub-sample} & \colhead{Number} & \colhead{Median Sample\tablenotemark{a}} & \colhead{Distribution} & \colhead{Distribution} & \colhead{Distribution} & \colhead{Distribution}\\
\colhead{Property}&\colhead{Description} & \colhead{of Obj.} & \colhead{S/N} & \colhead{Median\tablenotemark{b}} & \colhead{HIPR\tablenotemark{b}}&
\colhead{Mean\tablenotemark{b}} &\colhead{Std.\ Dev.\tablenotemark{b}}\\
\colhead{(1)} &
\colhead{(2)} &
\colhead{(3)} &
\colhead{(4)} &
\colhead{(5)} &
\colhead{(6)} &
\colhead{(7)} &
\colhead{(8)} 
}

\startdata
$\Delta$FWHM					& S/N$>$10	& 122	& 13.9	& -115 (-0.028)	& 438	& $-307$ (-0.085) 	& 725	\\ 
$\Delta$FWHM					& All			& 482	& 5.5		& -384 (-0.091)	& 628	& $-541$ (-0.164)	& 995	\\
$\Delta$FWHM					& S/N$<$10	& 360	& 4.1		& -471 (-0.125)	& 718	& $-629$ (-0.195)	& 1075	\\
\vspace{0.2cm}
$\Delta$FWHM					& S/N$<$5	& 218	& 2.9		& -513 (-0.142)	& 811	& $-718$ (-0.236)	& 1242	\\
$\Delta\sigma_l$				& S/N$>$10	& 122	& 13.9	& 132 (0.033)	& 547	& 117 (0.014)		& 862	\\ 
$\Delta\sigma_l$				& All			& 482	& 5.5		& 257 (0.073)	& 607	& 289 (0.059)		& 768	\\
$\Delta\sigma_l$				& S/N$<$10	& 360	& 4.1		& 294 (0.081)	& 613	& 347 (0.073)		& 725	\\
\vspace{0.2cm}
$\Delta\sigma_l$				& S/N$<$5	& 218	& 2.9		& 381 (0.104)	& 638	& 403 (0.086)		& 708	\\
$\Delta$MAD					& S/N$>$10	& 122	& 13.9	& 94 (0.029) 	& 309  	& 31 (-0.001)  		& 623  \\ 
$\Delta$MAD					& All			& 482	& 5.5		& 135 (0.046) 	& 363	& 133 (0.034) 		& 525  \\
$\Delta$MAD					& S/N$<$10	& 360	& 4.1		& 147 (0.053)  	& 384  	& 166 (0.045) 		& 484  \\
\vspace{0.2cm}
$\Delta$MAD					& S/N$<$5	& 218	& 2.9		& 188 (0.072) 	& 408  	& 187 (0.048) 		& 484  \\
$\Delta$EW					& S/N$>$10	& 122	& 13.9	& 3 			& 10 		& 2  				& 16  \\ 
$\Delta$EW					& All			& 482	& 5.5		& 9 			& 17 		& 13  			& 28  \\
$\Delta$EW					& S/N$<$10	& 360	& 4.1		& 11 	 		& 19  	& 17  			& 30  \\
\vspace{0.2cm}
$\Delta$EW					& S/N$<$5	& 218	& 2.9		& 17 			& 27  	& 23  			& 34  \\
$\Delta$Centroid				& S/N$>$10	& 122	& 13.9	& 0.19  		& 3.22  	& 6.78  			& 61.36  \\ 
$\Delta$Centroid				& All			& 482	& 5.5		& -0.03  		& 2.99  	& 0.10  			& 4.44  \\
$\Delta$Centroid				& S/N$<$10	& 360	& 4.1		& -0.20  		& 3.22  	& -0.28  			& 4.62  \\
\vspace{0.2cm}
$\Delta$Centroid				& S/N$<$5	& 218	& 2.9		& -0.20  		& 2.99  	& -0.49  			& 4.94  \\
$\Delta$Kurtosis				& S/N$>$10	& 122	& 13.9	& 0.32  		& 0.52  	& 0.39  			& 0.59  \\ 
$\Delta$Kurtosis				& All			& 482	& 5.5		& 0.52  		& 0.59  	& 0.65  			& 0.73  \\
$\Delta$Kurtosis				& S/N$<$10	& 360	& 4.1		& 0.63  		& 0.61  	& 0.75  			& 0.75  \\
\vspace{0.2cm}
$\Delta$Kurtosis				& S/N$<$5	& 218	& 2.9		& 0.77  		& 0.62  	& 0.89  			& 0.79  \\
$\Delta$Shape (FWHM/$\sigma_l$)	& S/N$>$10	& 122	& 13.9	& -0.07  		& 0.26  	& -0.12  			& 0.31  \\ 
$\Delta$Shape (FWHM/$\sigma_l$)	& All			& 482	& 5.5		& -0.22  		& 0.31  	& -0.23  			& 0.34  \\
$\Delta$Shape (FWHM/$\sigma_l$)	& S/N$<$10	& 360	& 4.1		& -0.29  		& 0.31  	& -0.27  			& 0.34  \\
$\Delta$Shape (FWHM/$\sigma_l$)	& S/N$<$5	& 218	& 2.9		& -0.35  		& 0.32  	& -0.31  			& 0.34 
\enddata

%

\tablenotetext{a}{The Median S/N is based on the distributions of SE spectra.  The S/N is measured per Angstrom, integrated over an emission-line-free continuum window, $\Delta W$, covering many resolution elements near restframe 1700\AA.}
\tablenotetext{b}{The median, HIPR, mean, and standard deviation (Std.\ Dev.) values are in units of \kms\ for the FWHM and $\sigma_l$ distributions and \AA\ for the EW and Centroid distributions.  The Kurtosis and Shape parameters are dimensionless. We also report the corresponding mean and medians of the fractional velocity distributions for the Coadded$-$SE Line Widths, $\Delta$FWHM/FWHM and $\Delta\sigma_{l}/\sigma_{l}$ in parentheses following each median and mean to be consistent with what is shown in Figure~\ref{F_widthscomparehist}.}

\label{T_widthdiffs}
\end{deluxetable}



\end{document}